\documentclass{article}
\usepackage[utf8]{inputenc}

\usepackage[english]{babel}

\usepackage[letterpaper,top=2cm,bottom=2cm,left=3cm,right=3cm,marginparwidth=1.75cm]{geometry}

\usepackage{amsmath}
\usepackage{graphicx}
\usepackage[colorlinks=true, allcolors=blue]{hyperref}
\usepackage[numbers]{natbib}
\usepackage{comment}
\usepackage{csquotes}

\title{One Venue, Two Conferences: The Separation of Chinese and American Citation Networks}
\author{
Bingchen Zhao$^*$ \\
  University of Edinburgh\\
  \texttt{bingchen.zhao@ed.ac.uk}
  \and
  Yuling Gu$^*$ \\
  Allen Institute for Artificial Intelligence\\
  \texttt{yulingg@allenai.org}
  \and
  Jessica Zosa Forde\\
  Brown University\\
  \texttt{jessica\_forde@brown.edu}
  \and
  Naomi Saphra\\
  New York University\\
  \texttt{nsaphra@nyu.edu}
}
\date{}

\begin{document}

\maketitle

\begin{abstract}
    At NeurIPS, American and Chinese institutions cite papers from each other's regions substantially less than they cite endogamously. We build a citation graph to quantify this divide, compare it to European connectivity, and discuss the causes and consequences of the separation.
\end{abstract}

\section{Introduction}

In recent years, the machine learning research landscape has been reshaped by the growth of Chinese AI research. China now consistently stands as the second-largest country in terms of total publications at NeurIPS, after the United States. In 2020, papers from Chinese institutions represented 13.6\% of all NeurIPS publications~\citep{ivanov_neurips_2020}.  The following year, this share increased to 17.5\%, representing a relative increase of 28.7\%~\citep{synced_neurips_2021}.\footnote{These numbers can be regarded as an underestimate of total share, as American multinational companies have been categorized as American regardless of the office location of the authors.}

In spite of China's position as a national AI powerhouse, collaborations between Chinese and American institutions are rarer than those between, say, American and Western European institutions \citep{rungta_geographic_2022}. Anecdotally, they also form distinct social groups at machine learning conferences: Chinese researchers often converse and take meals in separate groups from European and North American researchers. This divide goes beyond only social interactions; one prominent non-Chinese professor in an applied area of machine learning drew attention on Twitter when they advised students to avoid talks by Chinese authors, arguing that the presentations would be difficult to understand or of poor quality.\footnote{Out of a respect for privacy, we have kept this anecdote anonymous.} While it is true that many non-native anglophones find publicly speaking in English to be a challenge, avoiding talks by Chinese researchers may limit a conference attendee's exposure to new topics and ideas.
% Such a divide inevitably holds back a scientific community that could engage with each other's ideas on their own terms, and instead cites and engages endogamously.

In this work, we measure this divide between researchers in China and researchers in the United States.  Using NeurIPS citation data, we analyze the impact of work from US-based and China-based institutions.  We find that Chinese institutions under-cite work from the US and Europe, and both American and European institutions under-cite work from China.

\section{Citation networks}

How severe is this segregation between American and Chinese research communities? In order to measure the effect quantitatively, we collate citation data with institutions labeled by region. We consider the divisions between American, European, and Chinese researchers.  We find that European and American institutions are more closely linked through citations, and Chinese research is less linked to American and European research.

\subsection{Methods}
We compiled a citation graph by joining the citation data of the NeurIPS papers available from SemanticScholar\footnote{\url{https://www.semanticscholar.org/product/api}} with institutional information about authors from AMiner\footnote{\url{https://www.aminer.cn/oag-2-1}}. We first collected all paper titles from NeurIPS 2012 to 2021 from the NeurIPS website. We map the paper titles to their Semantic Scholar paper IDs using the Semantic Scholar Academic Graph (S2AG) API \citep{s2ag}. We manually searched for the unmatched papers, finding all but one in the Semantic Scholar database. For each of these papers, we then used the S2AG API to identify the authors, as well as authors of papers in its references.

\begin{figure*}[!h]
    \centering
    \includegraphics[width=0.4\linewidth]{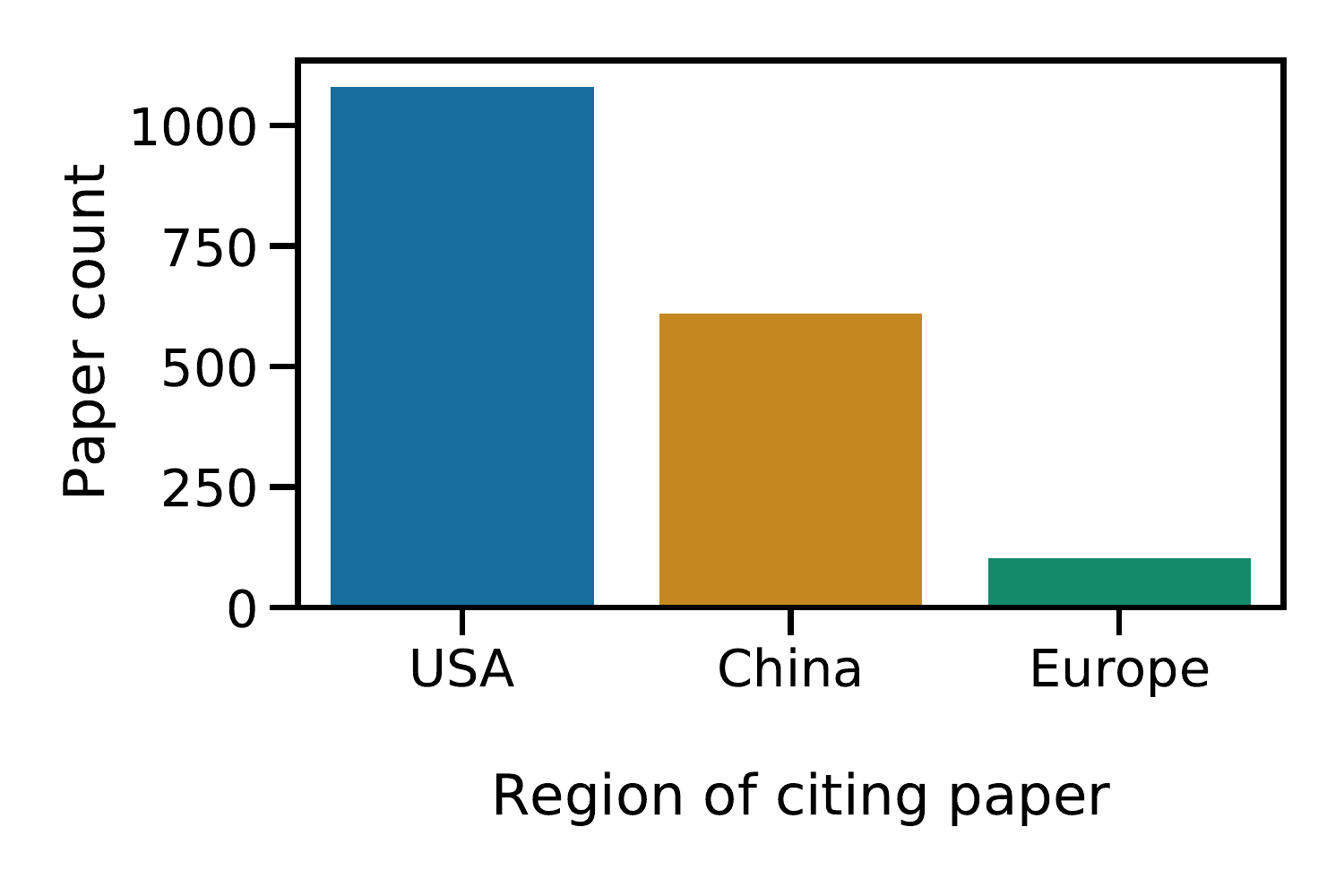}
    \includegraphics[width=0.4\linewidth]{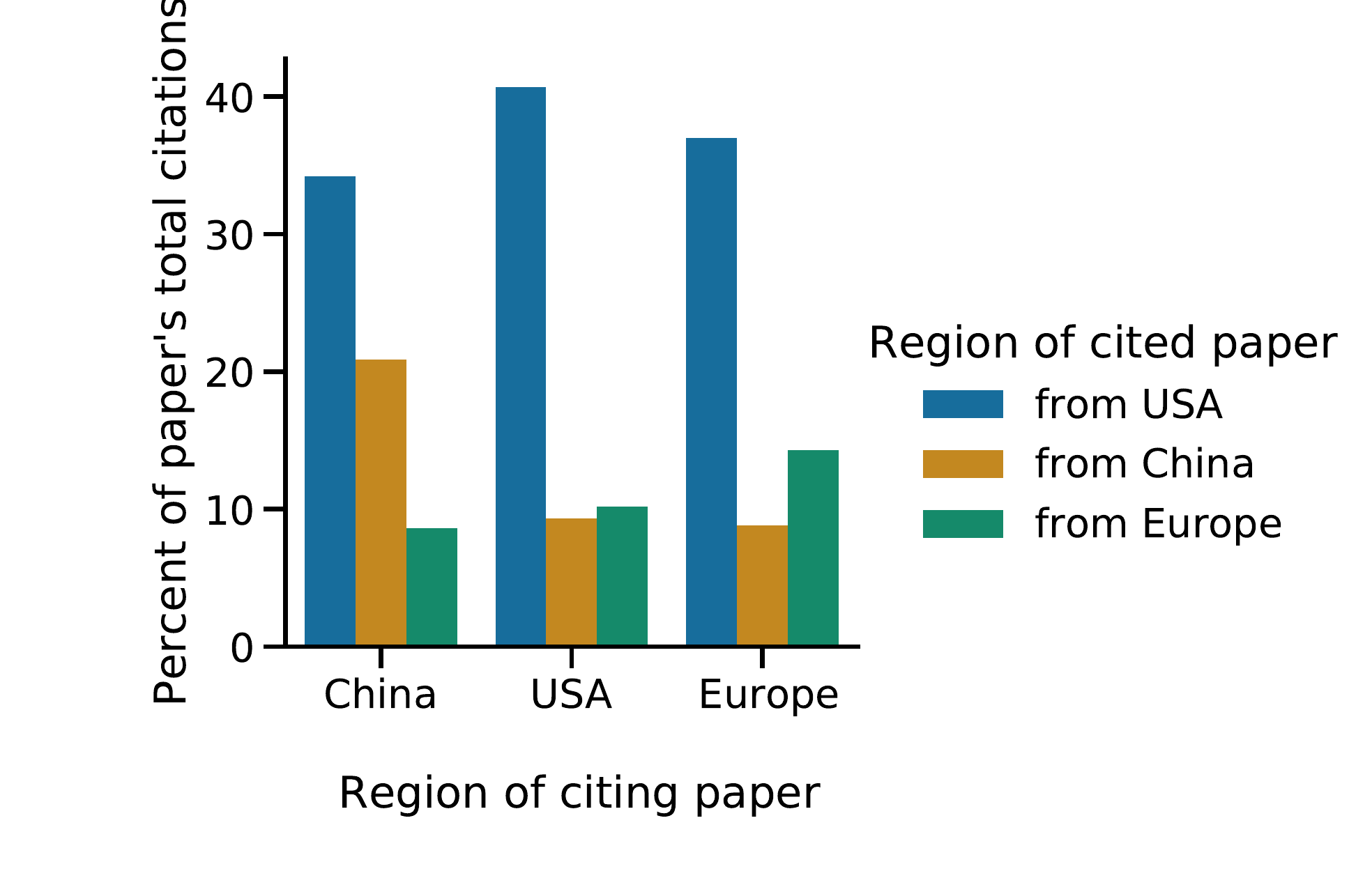} 
    \caption{Papers from Europe and the US cite one another more often than they cite papers from China, and papers from China are least likely to cite institutions in Europe and the US.}
    \label{fig:citations}
\end{figure*}

We then used AMiner to identify institutional information for each author. The 9460 NeurIPS papers have 135,941 authors in total, of which we found institutions for 83,515 authors (61\%). There are 4038 papers that have no authors on AMiner, so they are removed from the data. 
We then automatically marked institutes that included the name of a country, along with common cities and regions in China. We supplemented these automatic annotations with existing \footnote{\url{https://github.com/nd7141/icml2020/blob/master/university2.csv}} regional matchings and manually added 364 additional rules for regional matching.  Finally, we removed major multinational corporate labs (e.g., Google, Meta, Microsoft, Tencent, Alibaba, or Huawei), as they typically do not include information about the local office of the authors. Out of the remaining 5422 papers, we removed papers that weren't in one of the specified regions (China, US, Europe), or that included collaborators in multiple regions, leaving 1792 papers remaining. We then computed the average number and proportion of citations between papers from each region~(Figure~\ref{fig:citations}). 

\subsection{Results}

Based on these numbers, we see the extent to which American and Chinese papers fail to cite each other. While American papers make up 60\% of the dataset, they make up only 34\% of Chinese citations. American citations of Chinese papers show a more dramatic effect: while Chinese papers make up 34\% of the dataset, they are only 9\% of American citations. The effect is clear when we compare these numbers to American citations of European papers: even though our dataset of NeurIPS papers has six times as many Chinese papers as European papers, American institutions cite Chinese papers \textit{less} often than European papers. (Note that the cited non-NeurIPS papers can be older, and thus are less likely to be Chinese than the citing papers.)

We observe that each region tends to self-cite more frequently than they are otherwise cited: 21\% for China, 41\% for the USA, and 14\% for Europe. However, the separation between American and Chinese research communities is more pronounced than one might expect from typical regional preference. The American and European research communities share similar citation behavior, with few citations to Chinese papers, whereas Chinese institutions cite American and European papers less often than either other region does.

\begin{comment}
\begin{table}
    \centering
    \begin{tabular}{c|ccc}
        Region of citing paper: & China & USA & Europe \\
        \hline
        Paper count & 609 & 1081 & 102\\
        Mean citations across regions & 132.5  & 102.1 & 92.8 \\
        Mean Chinese citation count & 29.8  & 10.6 & 10.1  \\
        Mean American citation count & 43.5 & 39.5 & 32.8  \\
        Mean European citation count & 10.9 & 9.9  & 12.1   \\
    \end{tabular}
    \caption{Regional information for citations originating in a paper from China, the USA, or Europe.}
    \label{tab:citations}
\end{table}
\end{comment}

\subsection{Limitations}

The strength of our claims is subject to the influence of a number of factors we have not accounted for in this work. 

 First, while we consider the work from any university situated in the United States to be American, US-based labs may still have close links to China, which may lead us to underestimate the division between AI research in the US and China. For example, there exist labs in the US that are made up largely or entirely of Chinese international students. Similarly, Chinese international students who return to their home country may bring international connections to their alma mater and other institutions. We have not measured the degree to which such graduates change their citation patterns to focus on domestic papers, or whether they continue to widely cite American work. Additionally, our filter of multinational corporate labs may be incomplete. If these industry authors include companies with offices in both America and China, they may affect the reliability of our conclusions. 

There is another limitation that affects the confidence of our conclusions: the number of papers that have been excluded on the basis of authors not appearing in AMiner, a Chinese company. This is likely to result in there being more Chinese papers in the dataset than if we had full information about every paper. Overall, we discarded 43\% of papers due to a lack of author information, and this excluded set is likely to be a biased sample.

%For example, multi-object tracking is an active area of research in China, with large scale benchmarks such has Wang et al. \citep{wang2020panda}, which provides annotations of individual humans in each video. In the US, however, criticisms regarding of the misuse of biometric data \citep{bengio_2022} have led US researchers to avoid related tasks and datasets.  Similarly, the US tends to be heavily represented at fairness conferences such as FACCT, while representation from China remains limited \citep{Card_2022}.

\section{Consequences}

While American and Chinese researchers publish in the same venues, they represent two parallel communities, each producing work with limited impact on the other. To some degree this divide can be attributed to an interest in different topics, as cultural norms inform research priorities. For example, multi-object tracking is an active area of research in China, with large scale benchmarks such as Wang et al. \citep{wang2020panda}, which provides annotations of individual humans in each video. In North America, however, criticisms regarding of the misuse of biometric data \citep{bengio_2022} have led researchers to avoid related tasks and datasets.  Similarly, the US tends to be heavily represented at fairness conferences such as FACCT, while representation from China remains limited \citep{Card_2022}.

However, even abstract topics or architectures that are popular in China may not catch on in other regions. PCANet \citep{Chan2015PCANetAS}, an image classification architecture from a Singaporean research lab, has 1200 citations which are largely from Chinese or East Asian institutions. Deep Forests \citep{Zhou2017DeepFT}, which were developed at Nanjing University, garnered over 600 citations that are mostly Chinese. 

It is not only topics of research that are limited by a lack of exchange between regions. In recent years, the North American and European AI community has begun to engage in conversations and publishing research on the ethical considerations of AI. In accordance with these considerations, AI conferences have adopted systems for reviewers to flag ethical concerns and ask authors to write ethics statements or fill out checklists \citep{bengio_2022, rogers-etal-2021-just-think}. Engagement with researchers from China on these topics, however, remains limited. An example of this disconnect is the Provisional Draft of the NeurIPS Code of Ethics \citep{bengio_2022}.  At the time of initial publication, the authors' affiliations were with US universities, US-multinationals, and a university in Australia.  None of the authors were based in Asia.  This absence is despite the fact that observers have noted many similarities in ethics statements from Chinese based AI institutions \citep{Sheehan_2021, Roberts_Cowls_Morley_Taddeo_Wang_Floridi_2021}. 

Yet despite these similarities in ethics statements from US and Chinese researchers, specific disagreements in research practice still exist.
% However, international students are vulnerable to institutional mistreatment, regardless of the nationality of their PI, due to their reliance on their student visa from their university to remain in the country. While changing supervisors is difficult for many PhD students, these challenges are exacerbated for international students due to their immigration status. Moreover,  action may also be shielded by different cultural expectations of hierarchy in the advising relationship.\footnote{International student abuse, rarely discussed publicly, was brought to wider attention after University of Florida student Huixiang Chen took his own life. According to Chen's suicide note, his advisor, who was also Chinese, had pressured him to falsify results \citep{voice_hidden_2021}. } Regardless of the cause, a lab that is exclusively made up of Chinese students might be exposed more to Chinese and less to American work. In the more exploitative labs, furthermore, an abusive academic environment can limit interactions and networking with local research communities, leading to atypical citation patterns. Therefore, these few labs are likely to artificially \textit{inflate} the frequency with which American institutions cite Chinese work.
For example, Duke University stopped providing the Duke-MTMC dataset \citep{ristani2016MTMC} for studying object tracking across multiple cameras, because the researchers did not obtain consent from the students they collected images from \citep{tomasi_letter_2019}.
However, similar datasets like Market-1501  \citep{zheng2015scalable} from Tsinghua University are still being actively used \citep{he2021transreid,fu2022large}. 
Furthermore, researchers in China have continued to use the Duke dataset even after its removal \citep{wu2018cvpr_oneshot,LiTPAMI2019,zheng2017unlabeled}.
We can't know whether American researchers would have persuaded the authors with their arguments about the ethical hazards of such datasets, because they failed to engage the Chinese authors in these conversations.

The separation between these two communities has real impact on the individual researchers, the machine learning community as a whole, and potentially the societies affected by AI research. It is long past time that the AI community had a conversation about how to overcome this barrier.

% \subsection{Possible remedies}

% There are many ways to encourage collaboration, and therefore the diffusion of ideas, between two different groups. Workshops are an obvious route; international collaborations have developed from large summer programs like JSALT and the Les Houches summer school, but those programs remain largely American and European. \naomi{let's talk about what our or suggestions are going to be}
% \bingchen{What I can think of is in a form of ML Reproducibility Challenge, where people compete to implement the papers from a different region.}
% \yuling{effort to start viewing research produced by different groups equally, starting with judging research only by quality of the work its papers (rather than where the authors are from). Encourage more interaction between research groups (e.g. joint reading group between Chinese and US university labs to expose students to different papers, ideas and perspectives)}

% \begin{itemize}
%     \item something about internal presentations of work in Chinese
%     \item something about anglocentrism?
%     \item something about the consequences on machine learning research? How might this be shaping the topics people work on? maybe we should talk about the disconnection in terms of the American rise of ethics reviews as a source of conflict?
% \end{itemize}

\bibliography{main}
\bibliographystyle{abbrv}

\end{document}